\documentclass[preprint,12pt, a4paper]{elsarticle}

\usepackage{amssymb}

\usepackage{lineno}

\usepackage{url}
\usepackage[hidelinks]{hyperref}
\usepackage{listings}
\usepackage{tabularx}
\usepackage{dblfloatfix}
\usepackage[absolute]{textpos}

\usepackage{etoolbox}
\makeatletter
\patchcmd{\lsthk@TextStyle}{\let\lst@DefEsc\@empty}{}{}{\errmessage{failed to patch}}
\makeatother

\journal{SoftwareX}

\begin{document}

\begin{frontmatter}



\title{GeoRocket: A scalable and cloud-based data store for big geospatial files}


\author{Michel Kr{\"a}mer\corref{cor1}}
\ead{michel.kraemer@igd.fraunhofer.de}
\ead[url]{https://georocket.io}

\address{Fraunhofer Institute for Computer Graphics Research IGD, 64283 Darmstadt, Germany}

\begin{abstract}
We present GeoRocket, a software for the management of very large geospatial datasets in the cloud. GeoRocket employs a novel way to handle arbitrarily large datasets by splitting them into chunks that are processed individually. The software has a modern reactive architecture and makes use of existing services including Elasticsearch and storage back ends such as MongoDB or Amazon S3. GeoRocket is schema-agnostic and supports a wide range of heterogeneous geospatial file formats. It is also format-preserving and does not alter imported data in any way. The main benefits of GeoRocket are its performance, scalability, and usability, which make it suitable for a number of scientific and commercial use cases dealing with very high data volumes, complex datasets, and high velocity (Big Data). GeoRocket also provides many opportunities for further research in the area of geospatial data management.

\end{abstract}


\end{frontmatter}



\setlength{\TPHorizModule}{1mm}
\setlength{\TPVertModule}{1mm}
\begin{textblock}{200}(30,20)
  \small{$\copyright$ 2020. This manuscript version is made available under the CC-BY-NC-ND 4.0}
\end{textblock}
\begin{textblock}{200}(30,25)
  \small{license (\url{http://creativecommons.org/licenses/by-nc-nd/4.0/})}
\end{textblock}

\section{Motivation and significance}
\label{sec:introduction}

The global data volume is growing continuously. By the year 2025, it will have reached 163 zettabytes (or 163 trillion gigabytes) \cite{reinsel2017}. The main drivers of this data growth are mobile phones, autonomous cars, satellites, and other devices with built-in location sensors~\cite{goodchild2007}. Data collected by these devices can be located in time and place \cite{vatsavai2012} and is called \emph{spatiotemporal data} (or \emph{geospatial data}, \emph{geodata}). Kitchin \& McArdle recognise geospatial data as \emph{Big Data} characterised by its volume, variety, and velocity \cite{kitchin2016}. This means geospatial data sets are typically large, heterogeneous, and acquired in a short amount of time. Earth observation satellites, airborne laser scanners, and terrestrial mobile mapping systems, for example, record hundreds of thousands of samples per second \cite{cahalane2012} and produce a few GiB of data up to several TiB in a couple of hours \cite{paparoditis2012}. With the growing data volume, users face new challenges as their current computer systems lack storage space and computational power. At the same time, they require new software solutions capable of handling such data.

In our previous work, we investigated the possibilities of using the cloud and microservice architectures to process large amounts of heterogeneous geospatial data \cite{kraemer-2018,kraemer-senner-2015}. We focused on use cases from various domains such as land management, urban planning, and marine applications \cite{boehm-bredif-gierlinger-kraemer-lindenbergh-liu-michel-sirmacek-2016,iqmulus_d_1_2_3} where we could show that geospatial data can be of great value given there is sufficient computational power, enough storage resources, and suitable software.

To complement this, we now explore new ways to store, index, and query big geospatial data in a scalable, efficient, and inexpensive manner. We developed a novel software solution called \emph{GeoRocket} enabling users to store large amounts of geospatial vector data and to access, analyse, and share it in a distributed environment---in our case the cloud. The key properties of GeoRocket are its scalability, the indexing functionalities, as well as the modular architecture and lightweight interfaces. At the same time, it is schema-agnostic and format-preserving and provides users with a pragmatic way to store data.

With GeoRocket, we pursue a novel path that differentiates it from existing software solutions:

\begin{itemize}
\item \emph{PostGIS} \cite{postgis} is an extension to \emph{PostgreSQL} \cite{postgresql} that provides a low-level interface for application developers to store and analyse geospatial vector data in a traditional, relational database. In contrast, GeoRocket is a high-level data store that makes use of other storage technologies (see Section \ref{sec:description}). The geospatial entities in GeoRocket are semantic features and not geometries. GeoRocket is not a relational database. It also employs its own query language instead of SQL (see Section \ref{sec:sample-queries}).
\item \emph{GeoServer} \cite{geoserver} and \emph{Deegree} \cite{deegree} are storage solutions for geospatial data that have a long history. They have a monolithic architecture and use a traditional client/server approach. GeoRocket has a modern reactive and distributed architecture. It has been designed to run in the cloud and to harness the possibilities in terms of performance, scalability, and cost-effectiveness.
\item \emph{3DCityDB} \cite{3dcitydb} is a database that is specifically made to store CityGML files describing 3D city models \cite{ogccitygml}. CityGML is an application schema of the Geography Markup Language (GML) \cite{ogcgml}, which is itself based on XML. In contrast to 3DCityDB, GeoRocket supports multiple file formats and is schema-agnostic, so that it can handle CityGML, but also GML, or even arbitrary XML files. Besides, 3DCityDB has again a monolithic architecture.
\item \emph{rasdaman} \cite{rasdaman} is a storage and analytics solution for large geospatial raster data. It runs in a distributed environment and has been specifically designed for Big Data. The main difference to GeoRocket is the type of the data stored. GeoRocket supports vector data and rasdaman is made for raster data.
\item \emph{Cesium ion} \cite{cesiumion} is a commercial solution to host massive 3D datasets in the cloud and stream them efficiently for 3D web visualisation in the browser with the JavaScript framework Cesium \cite{cesiumjs}. Data uploaded to Cesium ion will be processed, optimized, and tiled to improve streaming and visualisation performance. The original data is not accessible. In contrast, GeoRocket allows access to the unmodified data. It also supports 2D as well as 3D data (albeit the supported file formats differ). It is a more generic solution that can be used in various applications, whereas Cesium ion focuses on 3D web visualisation. Both solutions complement each other and can be used in tandem.
\end{itemize}

The main contribution of this paper is the novel way to handle arbitrarily large data sets (see Section \ref{sec:description}). We describe GeoRocket's event-driven architecture and our approach to importing and indexing. We also show samples of GeoRocket's query language to demonstrate how GeoRocket can be used. Section \ref{sec:examples} describes an illustrative example where we used our software in a real-world application. GeoRocket provides potential for further research and commercial exploitation, which we discuss in Section \ref{sec:impact}. The paper finishes with conclusions in Section \ref{sec:conclusions}.

\section{Software description}
\label{sec:description}

\begin{figure}[t]
    \centering
    \includegraphics[width=.45\linewidth]{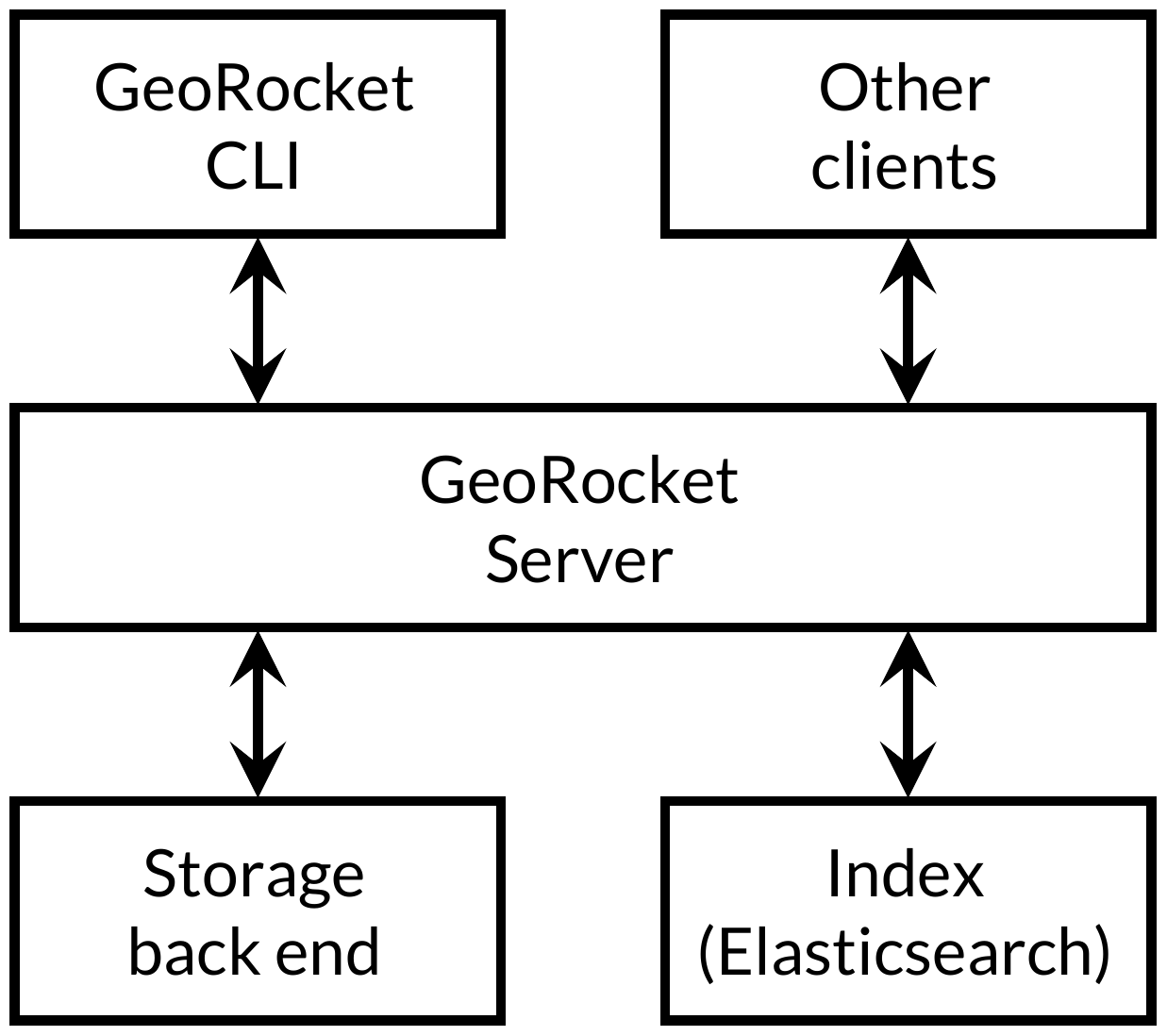}
    \caption{Overview of GeoRocket server, clients, and back-end services}
    \label{fig:overview}
\end{figure}

Figure \ref{fig:overview} shows a generic overview of a GeoRocket deployment. The main components are the GeoRocket server, the storage back end, and the index. The server is responsible for importing and exporting geospatial files. The actual data is kept in the storage back end. GeoRocket supports multiple back ends such as Amazon S3 \cite{amazons3}, MongoDB \cite{mongodb}, distributed file systems, or the local file system (typically used for testing purposes). In addition to the storage back end, GeoRocket keeps an inverted index about information found inside imported files. With this, users can search a large data set and extract the parts that are relevant to their use case. The index is maintained by the Open-Source framework Elasticsearch \cite{elasticsearch}. The processes of importing, indexing, and querying are described in Section~\ref{sec:architecture}.

In addition to the server, the storage back end, and the index, there is a command-line interface called GeoRocket CLI. It allows users to import and export files, as well as to manage \emph{tags} and \emph{properties} (see list of definitions below). The GeoRocket server also has an HTTP interface that can be used by other clients.

Before going into detail about the architecture of GeoRocket, we define a number of commonly used terms.

\begin{description}
\item[Chunk] A chunk represents a geospatial object (also called \emph{feature}) within an imported file. For example, in a CityGML file containing a 3D city model, a chunk represents a building (specified by a \lstinline!cityObjectMember! element). Analogously, in a GeoJSON file \cite{geojson}, a chunk is a feature in a feature collection. During import, GeoRocket splits geospatial files into individual chunks (see Section \ref{sec:architecture}) and saves them in its storage back end.
\item[Layer] A layer is a user-defined label (or folder, or directory) that can be used to structure a large collection of chunks in GeoRocket's storage back end. A chunk is always put into exactly one layer. If the user doesn't define one during import, the chunk will be put into the root layer called `\lstinline!/!'. Layers are structured hierarchically and parent layers always include all chunks of their sub-layers.
\item[Property] Properties are user-defined key-value pairs that can be attached to one or more chunks. Keys are unique.
\item[Tag] A tag is a user-defined label that can be attached freely to one or more chunks to structure data. Basically, a tag is a property with no value.
\item[Metadata] A metadata object includes user-defined tags and properties, as well as other automatically derived information (e.g. the imported file's spatial reference system).
\item[Indexed attribute] Indexed attributes are key-value pairs that GeoRocket detects during import. Unlike properties, they are not user-defined but directly extracted from the imported file (e.g. CityGML generic attributes or GeoJSON properties).
\end{description}

Note that chunks, layers, and indexed attributes are immutable. If a geospatial feature should be changed---i.e. if its attributes or geometry should be modified or if it should be moved from one layer to another---the feature has to be deleted and a modified one has to be imported again. User-defined metadata such as properties and tags, however, can be changed later.

This is also one of the reasons why we developed a new query language instead of using SQL. Joins and updates would be too complex or impossible to implement, in particular since GeoRocket is no relational database as described above.

\subsection{Software Architecture}
\label{sec:architecture}

GeoRocket has been implemented with Vert.x, an Open-Source toolkit for building reactive applications \cite{vertx}. Its architecture consists of so-called \emph{verticles}, which are independent components that communicate with each other by sending messages through an event bus. The software design adheres to the reactive manifesto \cite{reactive-manifesto}. GeoRocket is responsive, resilient, elastic, and message-driven. That means it is able to respond in a timely manner even under high load, and it provides good scalability in terms of data volume and number of users/parallel requests. At the same time it is fault-tolerant and can quickly recover from failures. Responsiveness, scalability, and elasticity are the result of both the event-driven architecture design based on Vert.x and the novel approach to importing and indexing. Fault-tolerance was implemented with patterns such as isolation, asynchronous timeouts, fail-fast, and retries. We refer to Nygard for more information on this topic~\cite{nygard2007}.

Note that these properties allow GeoRocket to be deployed to the cloud and to make use of the benefits offered by it. Individual verticles (or multiple instances of GeoRocket) can be deployed to distributed virtual machines (or even containers) in the cloud to achieve high performance, reliability, and scalability. The event-driven architecture with loosely coupled verticles (or instances) allows GeoRocket to scale elastically on demand, which can help optimise resource usage and ultimately reduce operating costs.

\begin{figure}[t]
    \centering
    \includegraphics[width=.47\linewidth]{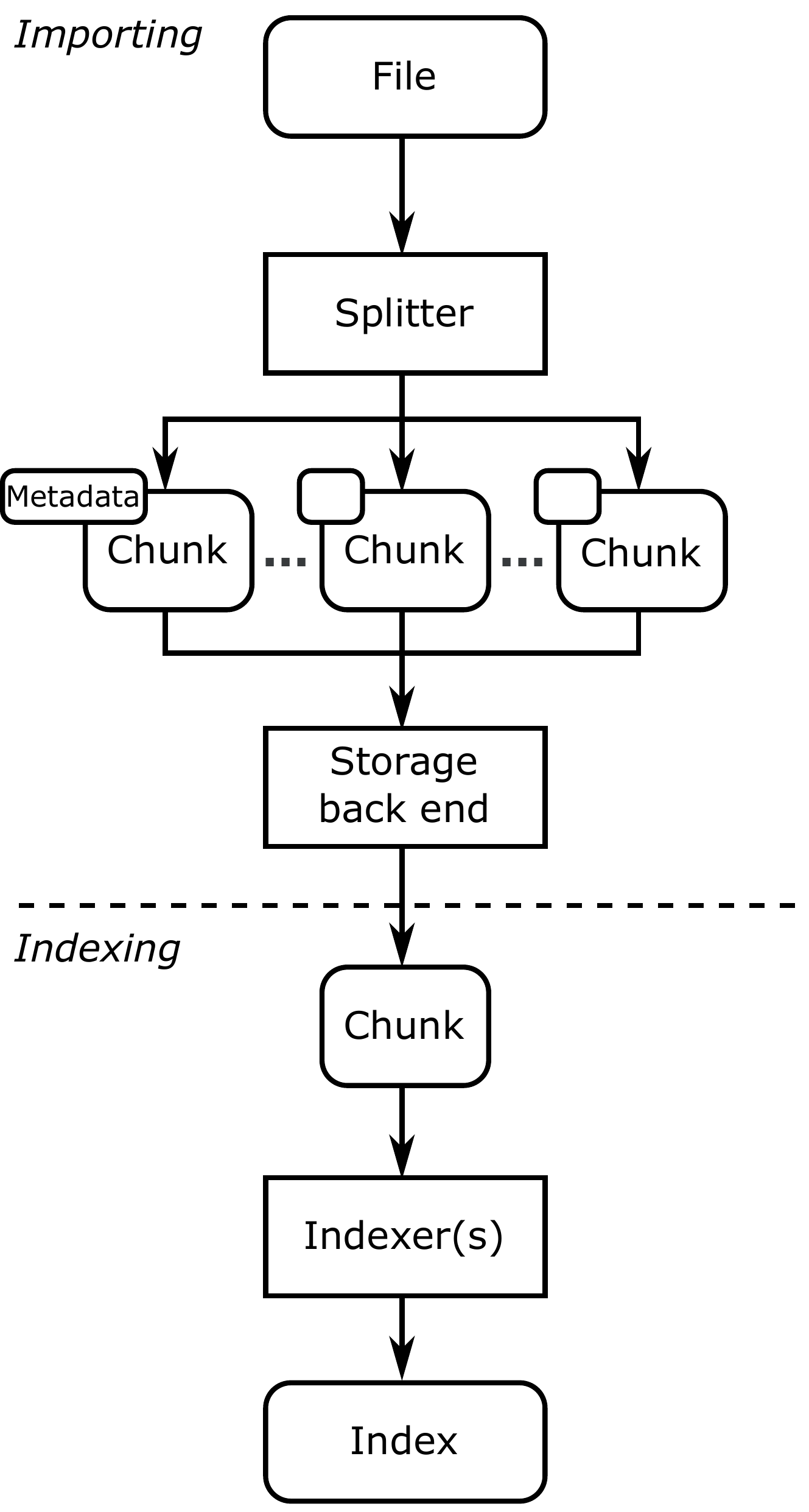}
    \caption{Importing and indexing a geospatial file: The file is split into chunks, which are in turn sent to the storage back end. After that, the indexing process runs asynchronously.}
    \label{fig:import}
\end{figure}

Figure \ref{fig:import} depicts the process of importing and indexing a geospatial file and how data flows between verticles. In order to be able to process arbitrary data volumes, GeoRocket uses a novel streaming approach that applies the divide-and-conquer paradigm. At the beginning, the geospatial file is divided into individual chunks by the \emph{Splitter} verticle. The Splitter also attaches user-defined metadata objects to each chunk. The chunks are saved in the configured storage back end. When all chunks have been written to the back end, the \emph{importing} process is finished.

As soon as the first chunk has been written, the \emph{indexing} process is started asynchronously. The \emph{Indexer} verticle reads every imported chunk from the storage back end and looks for known patterns such as attributes, geometries, or bounding boxes. For this, it uses lightweight stream-based parsing and regular expressions. This approach is faster and more scalable than loading the chunk completely into memory and interpreting it semantically. It also helps GeoRocket interpret geospatial data in a schema-agnostic manner. After parsing, the Indexer saves the extracted information into the index. Note that there can be more than one Indexer, each of them responsible for a certain kind of pattern. This allows GeoRocket to be extended with pluggable Indexers and to support indexing of heterogeneous data sets.

\begin{figure}[t]
    \centering
    \includegraphics[width=.6\linewidth]{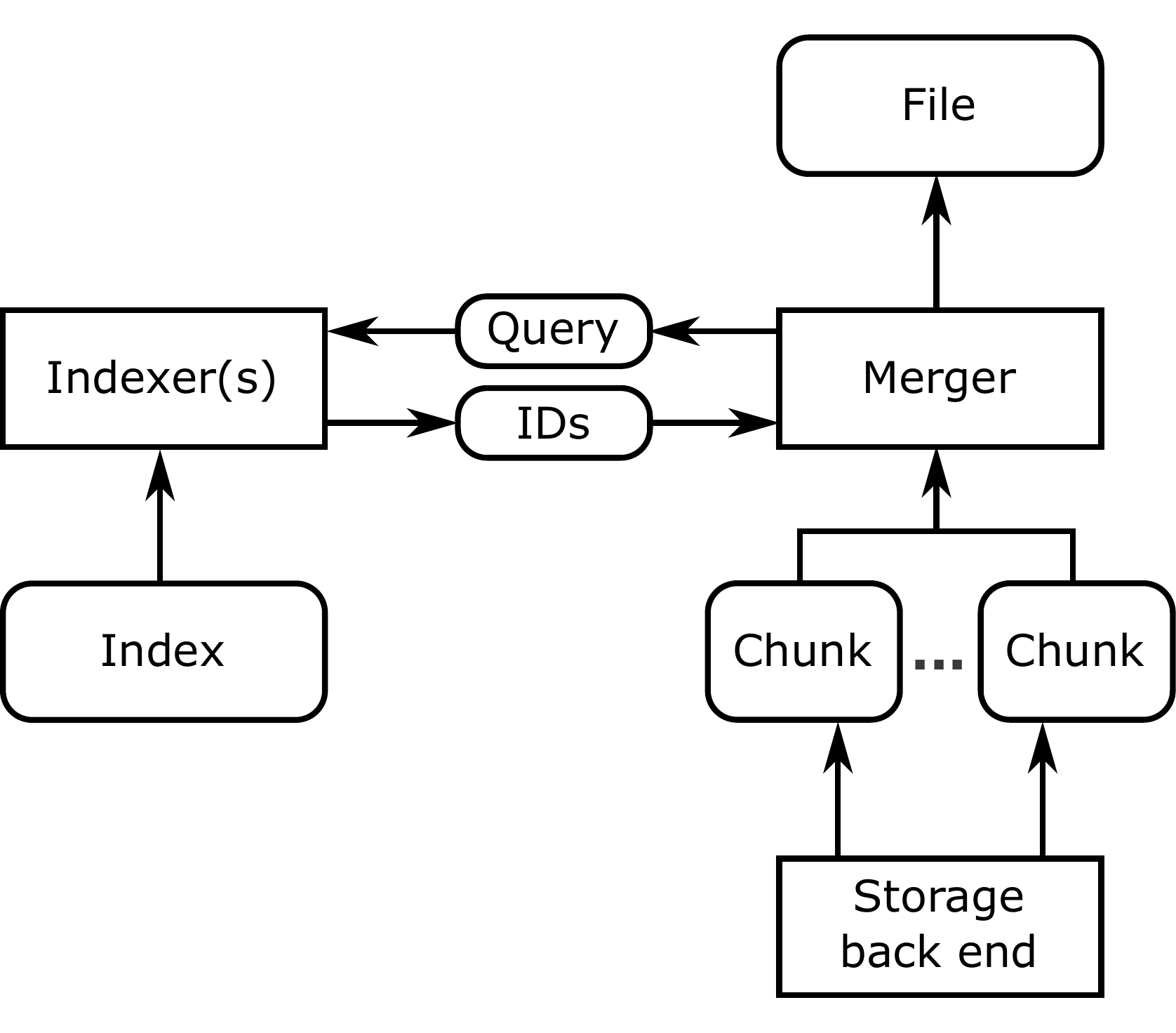}
    \caption{Exporting a file from GeoRocket: the merger retrieves chunk IDs from the indexer and merges matching chunks from the storage back end.}
    \label{fig:export}
\end{figure}

The process of querying and exporting files from GeoRocket is depicted in Figure \ref{fig:export}. The main component in this diagram is the \emph{Merger} verticle. It sends a query (see Section \ref{sec:sample-queries}) to the Indexers, which in turn search the index for chunk IDs matching the query's criteria. The chunk IDs are then sent back to the Merger, which in turn loads matching chunks from the storage back end. These chunks are joined to a valid output file that is finally rendered to the client.

\subsection{Software Functionalities}
\label{sec:functionalities}

From the software description above, we can derive the following key functionalities of our software:

\begin{itemize}
\item GeoRocket has been \emph{designed for the cloud}. It has a distributed architecture consisting of independent components (verticles) that can be deployed redundantly to achieve \emph{scalability}, \emph{performance}, and \emph{fault-tolerance}. This is in contrast to existing alternative software products that usually have a monolithic architecture.
\item It is backed by the Open-Source framework Elasticsearch that allows \emph{indexing and querying of very large data sets} in flexible ways and with high performance. Elasticsearch is itself designed to run in a distributed environment and fits well to the architecture of GeoRocket.
\item GeoRocket is \emph{schema-agnostic}, which means it does not require specific data schemas to work properly. Instead, it tries to identify common patterns in large heterogeneous data sets and uses the extracted information for indexing.
\item Since GeoRocket's data store is immutable, the software is \emph{format-preserving}. This means, every imported file can later, during export, be reconstructed as it was (apart from possible minor whitespace changes between chunks).
\end{itemize}

\subsection{Sample queries}
\label{sec:sample-queries}

In this section, we demonstrate how the query language of GeoRocket can be used to retrieve data. The structure of the language is lightweight. It consists of terms, logical operators, and comparison operators. Terms can be simple strings, dates, or bounding boxes (spatial areas defined by four numbers minimum X, minimum Y, maximum X, and maximum Y).
You can also use the logical operators \lstinline!AND!, \lstinline!OR!, and \lstinline!NOT!. The following example retrieves all chunks located inside the given bounding box \emph{and} containing the string \lstinline!Berlin! (e.g. as a value in one of the indexed attributes or properties) or that are labelled with the tag \lstinline!Berlin!:

\begin{lstlisting}
AND(13.378,52.515,13.380,52.517 Berlin)
\end{lstlisting}

You can also use comparison operators to constrain a term to a certain indexed attribute or property. The following complex example combines the logical operator \lstinline!AND! with the comparison operators \lstinline!EQ! (equals) and \lstinline!GTE! (greater than or equal to) to search for chunks that (a) lie inside a given bounding box, (b) whose indexed attribute or property \lstinline!name! equals \lstinline!Berlin!, and (c) whose indexed attribute or property \lstinline!importedDate! is greater than or equal to \lstinline!2018-02-13! (i.e. chunks that have been imported on or after this date):

\begin{lstlisting}
AND(13.378,52.515,13.380,52.517 EQ(name Berlin)
    GTE(importedDate 2018-02-13))
\end{lstlisting}

Note that, in this example, \lstinline!name! and \lstinline!importedDate! are either user-defined (if they are properties) or their existence depends on the imported data (in case they are indexed attributes). They are not created by default by GeoRocket.

A more detailed description of GeoRocket's query language can be found in the user documentation \cite{georocket-user-docs}.

\section{Illustrative Example}
\label{sec:examples}

In this section, we describe a real-world use case that demonstrates how GeoRocket can be used to store a very large geospatial dataset in a public cloud and to keep it up to date. The use case involves a data set of 3D building models provided by the German federal state of North Rhine-Westphalia (Land NRW). The dataset is in the CityGML format (Level of Detail 2) and is licensed under the dl-de/by-2-0 (Datenlizenz Deutschland - Namensnennung - Version 2.0, \href{http://www.govdata.de/dl-de/by-2-0}{www.govdata.de/dl-de/by-2-0}). It can be downloaded from \href{https://www.opengeodata.nrw.de/produkte/geobasis/3d-gm/3d-gm_lod2/}{www.opengeodata.nrw.de/produkte/geobasis/3d-gm/3d-gm\_lod2/}.

In order to demonstrate how this dataset can be kept up to date, we set up a GeoRocket cluster using \emph{Amazon Web Services (AWS)}. Table \ref{tab:georocket-aws} shows the EC2 instances we created and their configuration.

\begin{table}[!h]
\begin{tabularx}{\linewidth}{|X|l|c|r|r|}
\hline
\textbf{Description} & \textbf{Type} & \textbf{vCPUs} & \textbf{RAM} & \textbf{Volume size} \\
\hline
1 $\times$ GeoRocket 1.3.0 & c5.xlarge & 4 & 8 GiB & 40 GiB \\
\hline
3 $\times$ Elasticsearch 6.4.0 & m5.2xlarge & 8 & 32 GiB & 100 GiB \\
\hline
1 $\times$ MongoDB 4.0.2 & m5.large & 2 & 8 GiB & 100 GiB \\
\hline
\end{tabularx}
\caption{Instances of our GeoRocket cluster on AWS}
\label{tab:georocket-aws}
\end{table}

Our cluster consisted of five instances (1 for GeoRocket, 3 for Elasticsearch, and 1 for MongoDB) running in the AWS region eu-central-1b (Frankfurt). The volumes mounted into the instances were SSDs provided by the Amazon Elastic Block Store (EBS). All instances were running the Ubuntu 16.04 LTS AMI (Amazon Machine Image). We deployed and provisioned them using the Infrastructure-as-Code (IoC) tool Terraform \cite{terraform}.

After setting up the cluster, we imported the complete dataset with GeoRocket's command-line application (CLI). The dataset had a total size of 224.3 GB split up into 35,022 files. Since the CLI uses GZIP compression during upload, only 23.7 GB had to be transferred. We also recorded the space usage on our EC2 instances. The MongoDB database was 69.7 GiB large. The size of the sharded Elasticsearch index was 36.7 GiB. By default, Elasticsearch creates one replica of each index, so the total size of the Elasticsearch storage was 73.4 GiB distributed over the three EC2 instances. MongoDB and Elasticsearch use Snappy and LZ4 compression respectively, which is the reason why the used space was lower than the total size of our dataset. Both MongoDB and Elasticsearch contained entries for 10,529,668 chunks, which means there were this many geospatial objects in the dataset.

In order to demonstrate how such a large dataset can be managed with GeoRocket, we performed a workflow that is realistic and regularly happens in this or a similar way in municipalities or federal agencies. The dataset contained buildings in level of detail 2 (LoD2), which means they were only represented by wall and roof geometries. Suppose the dataset should be updated and more detailed building models should be added: for the purpose of city marketing, the LoD2 models of the popular shopping street `Schildergasse' in Cologne should be replaced by highly detailed geometries. Further suppose that old objects should not be removed from the dataset but kept for historical reasons.

First, we used the command-line application to mark the buildings in the Schildergasse in Cologne as outdated by adding a property \lstinline!deleted! denoting the date when the buildings were replaced. Since the dataset contained xAL 2.0 addresses \cite{xal}, we were able to use the terms \lstinline!Schildergasse! and \lstinline[escapechar=$]!K${\"o}$ln! (German for Cologne) in our command.\footnote{If you run the CLI on a Unix shell such as bash, you need to escape parentheses with backslashes. The Windows Command Prompt, on the other hand, does not require them. For the sake of readability, we omitted the backslashes here.}

\begin{lstlisting}[escapeinside=``]
georocket property set -props deleted:2018-09-13 \
  AND(Schildergasse K`{\"o}`ln)
\end{lstlisting}

We then imported the new buildings:

\begin{lstlisting}[escapeinside=``]
georocket import Schildergasse_update.gml
\end{lstlisting}

After this, we were able to download the complete city model of Cologne excluding the old models of the Schildergasse with the following query:

\begin{lstlisting}[escapeinside=``]
georocket search AND(NOT(LTE(deleted 2018-09-13)) K`{\"o}`ln)
\end{lstlisting}

This query matches all objects from Cologne but not those that have a \lstinline!deleted! property whose value is less than or equal to \lstinline!2018-09-13!.

All operations performed very fast. Setting the property was finished in a few milliseconds. The new file was only a few MiB large and importing it was a matter of seconds. The latency for downloading the complete city model of Cologne was again very low (a few milliseconds).

If the municipality or federal agency wanted to actually delete old data from the dataset on a regular basis for the sake of housekeeping (e.g. at the end of every year), they could use the following command:

\begin{lstlisting}[escapeinside=``]
georocket delete LT(deleted 2018)
\end{lstlisting}

This command would remove all objects from the dataset that have been marked as deleted in 2017 or earlier. GeoRocket is able to automatically parse dates in ISO format and compare their values accordingly.

\section{Impact}
\label{sec:impact}

As mentioned in Section \ref{sec:introduction}, geospatial data is increasingly becoming larger and more complex. Users are faced with new problems related to data volume and heterogeneity, as well as the speed with which data is acquired. One of the aims of developing GeoRocket was to make it possible to analyse and share such data by leveraging the possibilities of the cloud. This has opened up a number of new research directions and potential for changing the way users and companies work with Big Geo Data.

Firstly, we are currently working on developing novel visual analysis methods for large geospatial data sets. In the research project DataBio funded by the European Commission (grant agreement No 732064), we are using GeoRocket as a store for data from the agricultural domain. Based on this, we are developing a visual tool to interactively explore the data set and to perform analyses and aggregations.

We are also using GeoRocket in the area of Smart City Clouds where it enables users to access and share the large amounts of information collected in a Smart City for the first time for use cases such as urban planning or traffic management. In this respect, we have extended GeoRocket with means to encrypt data in the cloud while keeping the possibility to search it using Searchable Symmetric Encryption (SSE) \cite{hiemenz-kraemer-2018}. We have also explored the possibility to share geospatial data in a secure way in a Smart City Cloud for applications related to security \cite{kraemer-frese-2019}. Furthermore, we discussed the possibility to use GeoRocket in the area of processing of large geospatial data for use cases such as land monitoring or urban planning \cite[pp. 48--49 and 163--164]{kraemer-2018}.

Besides the research opportunities, we believe GeoRocket also benefits users and companies. As mentioned in Section \ref{sec:introduction}, there are existing products that can manage geospatial data, but they typically have a monolithic software architecture and are supposed to run in a traditional client/server setting. GeoRocket, on the other hand, has been designed to run in the cloud and to leverage its possibilities, not only in terms of performance and scalability but also cost-effectiveness. Deploying GeoRocket to the cloud can be much less expensive for users and companies than maintaining dedicated on-premise hardware. This particularly applies to public administrations or small and medium enterprises.

\section{Conclusions}
\label{sec:conclusions}

In this paper, we presented GeoRocket, a scalable and cloud-based data store for geospatial files. We compared it to existing products and described its architecture and query language. We also presented an illustrative example showing how GeoRocket can be used in a real-world application. Finally, we discussed the impact of our software with regards to opportunities for scientific research and commercial exploitation.

There is an ongoing paradigm shift in Computer Sciences towards Cloud Computing. This particularly applies to Geoinformatics, which has only started to make successful use of the cloud. GeoRocket is one of the first applications that is specifically designed to manage geospatial data and to run in the cloud. In this paper, we were able to show that our novel approach to data handling based on splitting files into chunks and indexing them individually has many benefits regarding performance, scalability, and usability. Since GeoRocket is schema-agnostic, it supports a wide range of geospatial datasets and can be used in multiple applications. It is also format-preserving and avoids information loss that typically happens when you have to transform data between different models. These properties make GeoRocket superior to existing solutions.

The illustrative example presented in the paper only scratches the surface of what is possible with GeoRocket. In Section \ref{sec:impact}, we already mentioned briefly that we are also working on visual analysis methods and secure data storage. In the future, we will focus on novel approaches to increase the use of the large amounts of data managed by GeoRocket through Big Data methods and Visual Analytics. We will also further improve the performance of GeoRocket and explore its use for time series and other spatiotemporal data.





\bibliographystyle{elsarticle-num-custom-webpages} 
\bibliography{paper.bib}






\pagebreak

\section*{Required Metadata}

\section*{Current code version}

\begin{table}[!h]
\begin{tabular}{|l|p{6.5cm}|p{6.5cm}|}
\hline
\textbf{Nr.} & \textbf{Code metadata description} & \textbf{Please fill in this column} \\
\hline
C1 & Current code version & Git SHA ace352a \\
\hline
C2 & Permanent link to code/repository used for this code version & \url{https://github.com/georocket/georocket/tree/v1.3.0} \\
\hline
C3 & Legal Code License   & Apache-2.0 \\
\hline
C4 & Code versioning system used & git \\
\hline
C5 & Software code languages, tools, and services used & Java, Vert.x, Elasticsearch \\
\hline
C6 & Compilation requirements, operating environments \& dependencies & JDK 8 \\
\hline
C7 & If available Link to developer documentation/manual & \url{https://georocket.io/docs/} \\
\hline
C8 & Support email for questions & \href{mailto:michel.kraemer@igd.fraunhofer.de}{\nolinkurl{michel.kraemer@igd.fraunhofer.de}} \\
\hline
\end{tabular}
\caption{Code metadata (mandatory)}
\end{table}

\vspace{-1.8em}\section*{Current executable software version}

\begin{table}[!h]
\begin{tabular}{|l|p{6.5cm}|p{6.5cm}|}
\hline
\textbf{Nr.} & \textbf{(Executable) software metadata description} & \textbf{Please fill in this column} \\
\hline
S1 & Current software version & 1.3.0 \\
\hline
S2 & Permanent link to executables of this version  & \url{https://github.com/georocket/georocket/releases/tag/v1.3.0} \\
\hline
S3 & Legal Software License & Apache-2.0 \\
\hline
S4 & Computing platforms/Operating Systems & Linux, macOS, Windows \\
\hline
S5 & Installation requirements \& dependencies & Java 8 \\
\hline
S6 & If available, link to user manual - if formally published include a reference to the publication in the reference list & \url{https://georocket.io/docs/user-documentation/1.3.0} \\
\hline
S7 & Support email for questions & \href{mailto:michel.kraemer@igd.fraunhofer.de}{\nolinkurl{michel.kraemer@igd.fraunhofer.de}} \\
\hline
\end{tabular}
\caption{Software metadata (optional)}
\end{table}

\end{document}